\documentclass[twocolumn,journal]{IEEEtran}
\usepackage[dvips]{graphicx}
\usepackage[english]{babel}
\usepackage{algorithm}
\usepackage{algorithmic}
\usepackage{amsfonts}
\usepackage{booktabs}
\usepackage{multirow}
\usepackage{stfloats}
\usepackage[numbers,sort&compress]{natbib}
\usepackage{amsmath}
\usepackage{amssymb}
\usepackage{array}
\usepackage{url}
\usepackage{color}
\usepackage{caption}

\usepackage{epsfig,latexsym}
\usepackage{flushend}

\title{Structured Satellite-UAV-Terrestrial Networks for 6G Internet of Things}

\author{
\IEEEauthorblockN{Wei~Feng, \IEEEmembership{Senior Member, IEEE}, Yanmin~Wang, Yunfei~Chen, \IEEEmembership{Senior Member, IEEE}, Ning~Ge, \IEEEmembership{Member, IEEE}, and Cheng-Xiang~Wang, \IEEEmembership{Fellow, IEEE}}\\
\thanks{W.~Feng and N. Ge are with Tsinghua University, China; Y.~Wang is with Minzu University of China, China; Y. Chen is with the University of Durham, U.K.; C.-X.~Wang is with Southeast University, and also with Purple Mountain Laboratories, China.}
}

\begin{document}
\maketitle

\begin{abstract}
The upcoming sixth generation (6G) wireless communication network is envisioned to
cover space, air, and maritime areas, in addition to urban-centered terrestrial coverage by the fifth generation (5G) network, to support intelligent Internet of Things (IoT). Towards this end, we investigate structured integration of satellites, unmanned aerial vehicles (UAVs), and terrestrial networks, aiming to serve future universal IoT possibly with a massive number of devices in the coverage holes of current 5G. The hybrid satellite-UAV-terrestrial network usually leads to high system complexity, due to the heterogeneity and dynamics of space/air/ground links. With a systematic thinking, we propose to create and exploit hierarchies for the integrated network. Four basic structures are discussed by learning from the synergies in our human body. To orchestrate multiple heterogeneous basic structures, we further propose a process-oriented on-demand coverage method, which characterizes the system behavior as a series of events over time and is able to tackle the system complexity elaborately. We also outline open issues for promoting the agility and intelligence of structured satellite-UAV-terrestrial networks in the making.
\end{abstract}

\IEEEpeerreviewmaketitle

\section{Introduction}
With the wide deployment of fifth generation (5G) networks, more and more research attention has been attracted to future sixth generation (6G) networks from both academia and industry~\cite{r1,r2}. Despite of existing fruitful efforts, it remains open to define the whole picture of 6G.
Nevertheless, one thing is certain that more and more Internet of Things (IoT) devices will require 6G services in space, air, and maritime areas. Thus, 6G should be able to provide a high-performance universal coverage, rather than only focusing on the terrestrial area as current 5G networks.

As defined by ITU-R~\cite{r3}, key performance metrics of 5G include peak data rate, user experienced data rate, latency, mobility, connection density, energy efficiency, spectrum efficiency, and area traffic capacity. It is noted that the coverage metric has been ignored for 5G, as the cellular architecture still dominates in 5G networks and it can offer desirable coverage if the base station (BS) sites are sufficient and properly planned. In the urban area,
this usually leads to an ultra-dense network, which costs a lot but essentially ensures the profit for 5G networks. 

When it comes to remote rural or maritime areas, things become extremely challenging. On one hand, the available BS sites are often quite limited due to geographical limitations. Taking the maritime scenario as an example, the BSs can only be deployed along the shore, or on some well-developed islands. Therefore, the cellular architecture might be invalidated, inevitably resulting in poor coverage. On the other hand, the sparse and uneven spatial distribution of users renders it wasteful to establish an ultra-dense cellular network in rural settings. Actually, the deficiency of 5G coverage has been noticed in some previous research works. For instance, it was pointed out that the 5G cellular architecture has never been successful in bringing cost-effective service to the rural area~\cite{r4}. Likewise, the authors of~\cite{r5} focused on 5G in rural and low-income areas, and pointed out that 5G technologies are ``urban'' in their nature, which means that its high performance is achieved at the expense of extremely ``rich'' communication infrastructures.

Intuitively, to efficiently extend the coverage area of 5G networks to global coverage, non-terrestrial facilities such as satellites and unmanned aerial vehicles (UAVs) should be integrated~\cite{r6}. Moreover, non-cellular coverage methodology should be contrived to adapt to the sparse and uneven distribution of IoT devices outside the cellular coverage. These understandings inspire an agile coverage-oriented non-cellular hybrid satellite-UAV-terrestrial network. In this article, we summarize requirements and challenges of integrating satellites, UAVs and terrestrial networks. To address the problem of system complexity, we propose to create and exploit hierarchies of the integrated network by mimicking the synergetic behaviors of our human body. Four basic structures are accordingly presented. To orchestrate these basic structures, we propose a process-oriented on-demand coverage optimization method, which solves the complexity problem at a newly introduced mesoscopic scale, in addition to conventional micro and macro processing scales.  

\section{Use Cases and Challenges for Satellite-UAV-Terrestrial Networks}
In this section, we first summarize the use cases of 6G IoT, which require ubiquitous connectivity far beyond the capability of current 5G. Then, we discuss technical challenges for an efficient space-air-ground integration.

\begin{figure}[!t]
  \centering
  \includegraphics[width=2.9 in]{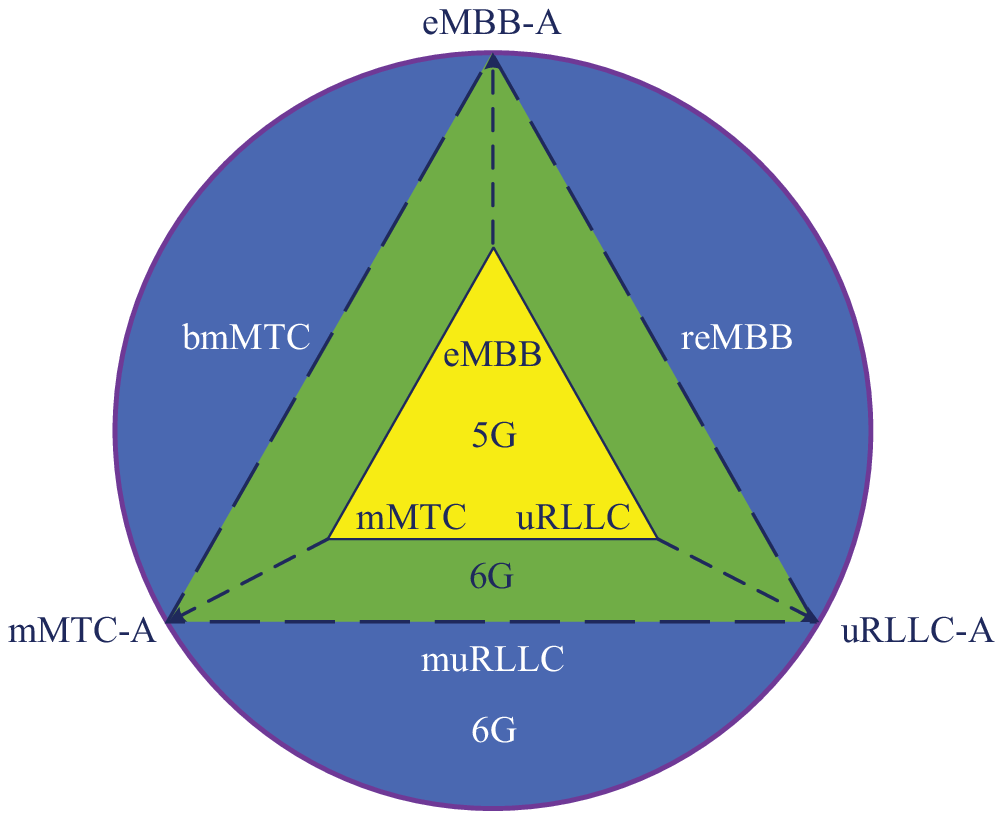}
  \caption{Illustration of use cases of 5G and 6G Internet of Things.}\label{fig1}
\end{figure}

\subsection{Use Cases}
In addition to the eight key performance metrics, ITU-R has also defined three use cases of 5G~\cite{r3}. They are enhanced mobile broadband (eMBB), massive machine type communications (mMTC), and ultra-reliable and low latency communications (uRLLC), as shown by the yellow triangle in Fig.~\ref{fig1}. Considering the ubiquitous feature of 6G, we extend these use cases to eMBB anywhere (eMBB-A), mMTC anywhere (mMTC-A), and uRLLC anywhere (uRLLC-A), as shown by the green triangle in Fig.~\ref{fig1}. Furthermore, these typical scenarios could be tightly coupled in practice, rendering a wide variety of their combinations, e.g., broadband mMTC (bmMTC), reliable eMBB (reMBB), massive uRLLC (muRLLC), and so on, as shown by the blue circle in Fig.~\ref{fig1}. In the following, we discuss these possible use cases in detail.

\textbf{eMBB-A}. The main target of eMBB-A is human being, which is the center of 2G/3G/4G/5G. In the 6G era, it remains one of the central missions to offer people with high-quality communication services with satisfying experiences. Different from eMBB of 5G, which requires high traffic capacity for hotspots but much lower data rate for wide area coverage, eMBB-A requires an always-online broadband service regardless of user locations. This is meaningful to support the ever-growing human-activity region, as well as some exploration activities such as scientific expeditions.

\textbf{mMTC-A}. The main target of mMTC-A is sensors, which typically transmit relatively small amount of non-delay-sensitive data. The difference between mMTC-A and mMTC in 5G~\cite{r3} is the coverage area. mMTC-A requires a universal coverage to support ubiquitous connections for any sensors, while mMTC of 5G typically depends on the cellular network, e.g., the narrowband IoT (NB-IoT) system, resulting in a limited coverage performance. mMTC-A is useful for resource detection in the remote rural area and environmental monitoring on the vast ocean.

\textbf{uRLLC-A}. The main target of uRLLC-A is robots, e.g., unmanned vehicles and automatic equipments for industrial manufacturing and unmanned operation. The main difference between uRLLC-A and uRLLC is their coverage area. Particularly, uRLLC is mostly performed in a local or pre-defined region. With the fast development of robots, uRLLC is needed anywhere, leading to uRLLC-A. In the 6G era, it can be envisioned to control a robot swarm anywhere in the world, relying on uRLLC-A.

\textbf{bmMTC}. The main target of bmMTC is enhanced sensors, which are massive as mMTC and also require broadband communications as eMBB.
For example, large numbers of high-definition cameras, which might be necessary for future Digital Twins and Metaverse, may motivate bmMTC. In summary, although both have massive devices, 5G's mMTC is for small data, whereas bmMTC is for big data, which could be crucial for 6G intelligent IoT.

\textbf{reMBB}. The main target of reMBB is intelligent robots. For example, a group of robots could be dispatched into a post-disaster area with unknown dangers. Some of them gather holographic multimedia information about the environment and send the data to other robots for swarm intelligence-empowered decision making. In this case, reMBB is required to guarantee both efficiency (like eMBB) and reliability (like uRLLC) of inter-robot communications.

\textbf{muRLLC}. The main target of muRLLC is also intelligent robots. Using the same example as above for reMBB, the number of robots can be very large for a vast region. All the robots should act in a coordinated manner, with frequent exchange of controlling signals.
These require muRLLC, so as to support massive users (like mMTC) with reliable and low-latency (like uRLLC) transmissions.

To support these use cases, one cannot rely only on the terrestrial network, as its coverage area is often limited due to geographical limitations. It is also infeasible to rely only on the satellite network, as its data rate is relatively low and its delay is often large due to long transmission distances. Therefore, it is necessary to integrate satellites, UAVs and terrestrial networks into one network, to support these applications in a synergetic fashion.

\subsection{Challenges}
\begin{figure}[!t]
  \centering
  \includegraphics[width=2.8 in]{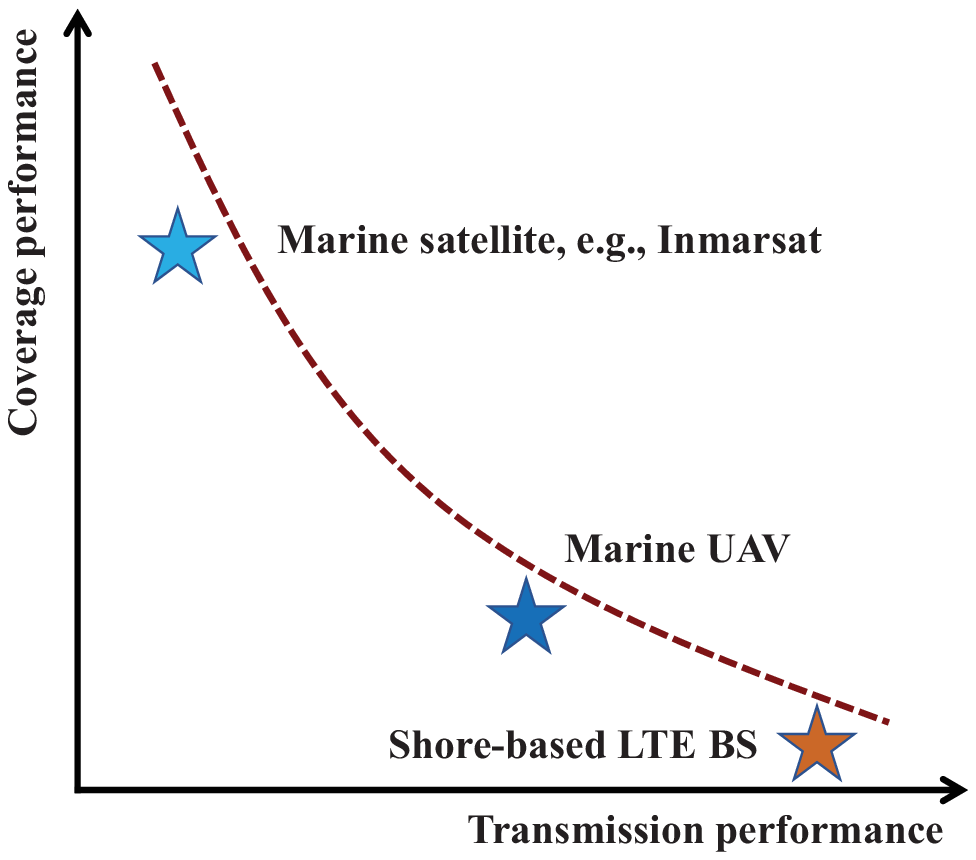}
  \caption{Illustration of the tradeoff between coverage and transmission performance: a maritime scenario example.}\label{fig2}
\end{figure}

In general, the above defined use cases are all quite challenging. For example, given total system cost, data rate and coverage often trade off each other for eMBB-A, and connection quantity and coverage trade off each other for mMTC-A. Likewise, latency/reliability and coverage for uRLLC-A, data rate and connection quantity for bmMTC, latency/reliability and data rate for reMBB, and connection quantity and latency/reliability for muRLLC trade off each other. We show the tradeoff between coverage and transmission performance in Fig.~\ref{fig2}, where the maritime scenario is taken as an example. In the figure, better transmission performance can be higher date rate, better reliability, or smaller latency. It is observed that the marine satellite is good at coverage, while the shore-based Long Term Evolution (LTE) BS is good at achieving a satisfactory transmission performance. The marine UAV lies in the middle, due to its controllable mobility. The dashed curve in the figure represents the fundamental tradeoff limit, which definitely exists and can be accurately depicted in a specific case with certain system settings. Taking the coverage distance-date rate pair as an example, longer distance usually renders lower rate, from the Shannon's capacity theory.

The inherent tradeoffs imply both challenges to realize high-performance wide-area communication services and necessity to integrate space-air-ground approaches. It also unveils the significant heterogeneity of space/air/ground links, which poses great challenges for efficient integration. We summarize the characteristics of satellites, UAVs, and terrestrial BSs as below.

\textbf{Satellite.} Although not restricted by geographical conditions, satellites move along fixed orbits. They are usually far away from ground IoT devices, resulting in relatively low data rate, large latency and wide coverage area. Satellites are generally expensive but they are not affected by terrestrial disasters, thus being quite valuable for emergency communications.

\textbf{UAV.} UAVs can be flexibly scheduled thanks to their controllable mobility. Nevertheless, weather conditions usually have a great impact on the deployment of UAVs. For example, above the ocean, only some specific type of UAVs, e.g., some oil-powered fixed-wing UAVs, can be adopted so as to tackle the harsh marine meteorological conditions. The transmission performance of UAVs is satisfactory as they can move as close as possible to targeted IoT devices. However, the schedule of UAVs should be carefully designed, which complicates the network management.

\textbf{Terrestrial BS.} It is the foundation of 2G/3G/4G/5G. Its deployment relies on geographical conditions, and its performance largely relies on its density. For urban areas, terrestrial BSs might be the best option for wireless coverage with high data rate and low latency. However, as the coverage range of a single BS is quite limited, it is expensive to adopt terrestrial BSs in the remote rural and maritime areas.

These different characteristics may further render intractable system complexity, in terms of e.g., dynamic non-cellular network topology and heterogeneous interfering links. This complexity comes from the huge heterogeneity of satellite, UAV and terrestrial links in terms of transmission rate, latency, coverage range, and so on~\cite{r7,r8}. It also comes from the non-cellular network topology, which dynamically changes with the mobility of satellites and UAVs, as well as being undecomposable in contrast to the tractable decomposability of cellular architecture. If
this complexity is not handled with care, its side effect, e.g., the overhead for acquiring channel state information (CSI) of each link~\cite{r7,r8}, may consume an unbearable amount of system resources, rendering failure of the integration in practice.

The above deduction follows the rule of \emph{limits to growth}~\cite{r9}, which actually has already been encountered in the area of wireless communications. A typical example is the massive MIMO technology, which is one of the most important enablers of 5G. Similar to integrating satellites and UAVs into terrestrial networks, a massive MIMO is formed by adding more antenna elements at the BS. Its \emph{limits to growth} was shown in~\cite{r10} as ``\emph{the number of terminals that can be simultaneously served is limited, not by the number of antennas, but rather by our inability to acquire channel-state information for an unlimited number of terminals}.'' It is just as the example in~\cite{r9}, i.e., the depletion of nonrenewable resources and the deteriorating environment could lead to the degradation of industrialization and the decrease of population, because these aspects are all interconnected in many ways within the huge complex system.

\section{Thinking in Systems and Structured Perspective}
In order to handle the enormous complexity in the integrated network, we leverage the basic principle of systematic thinking~\cite{r11}. In this section, we first explain the basic methodology, and then propose four basic structures through learning from the synergies in our human body, which is a full-fledged system handling complexities quite well.

\subsection{Creating and Exploiting Hierarchies}
We highlight the following excerpts from \emph{Meadows}'s book.

\begin{itemize}
  \item \emph{A system is more than the sum of its parts. It may exhibit adaptive, dynamic, goal-seeking, self-preserving,
and sometimes evolutionary behavior~\cite{r11}.}
  \item \emph{System structure is the source of system behavior. System behavior reveals itself as a series of events over time~\cite{r11}.}
  \item \emph{In the process of creating new structures and increasing complexity, one thing that a self-organizing system often generates is hierarchy~\cite{r11}.}
\end{itemize}

These rules are simple but insightful. They indicate that systematic thinking is useful for a better understanding of the complex behavior of systems, in addition to focusing on various part-wise design and analysis. We thus tend to combine \emph{Systems Theory} and \emph{Reductionism}, so as to design and operate a usually-complex satellite-UAV-terrestrial network, from a new structured perspective. Particularly, we are inspired to create and exploit hierarchies of the integrated network, so as to uncover the basic structures, i.e., the minimum units, for integration. Then, we would handle the system complexity via orchestrating multiple basic structures. Moreover, we notice that ``a series of events over time'' are usually tightly coupled within a system, and thus it is meaningful to have a process-oriented perspective for orchestrating heterogeneous basic structures. We then propose the process-oriented on-demand coverage regime, which will be described in Section IV.

\subsection{Learning from the Synergies in Our Human Body}
How to create and exploit hierarchies of the integrated network remains an open problem. We give a simple solution by learning from the synergetic behaviors in our human body.
Our body is a complex system, which handles various complexities quite efficiently using natural hierarchies.

\begin{figure}[!t]
  \centering
  \includegraphics[width=3.4 in]{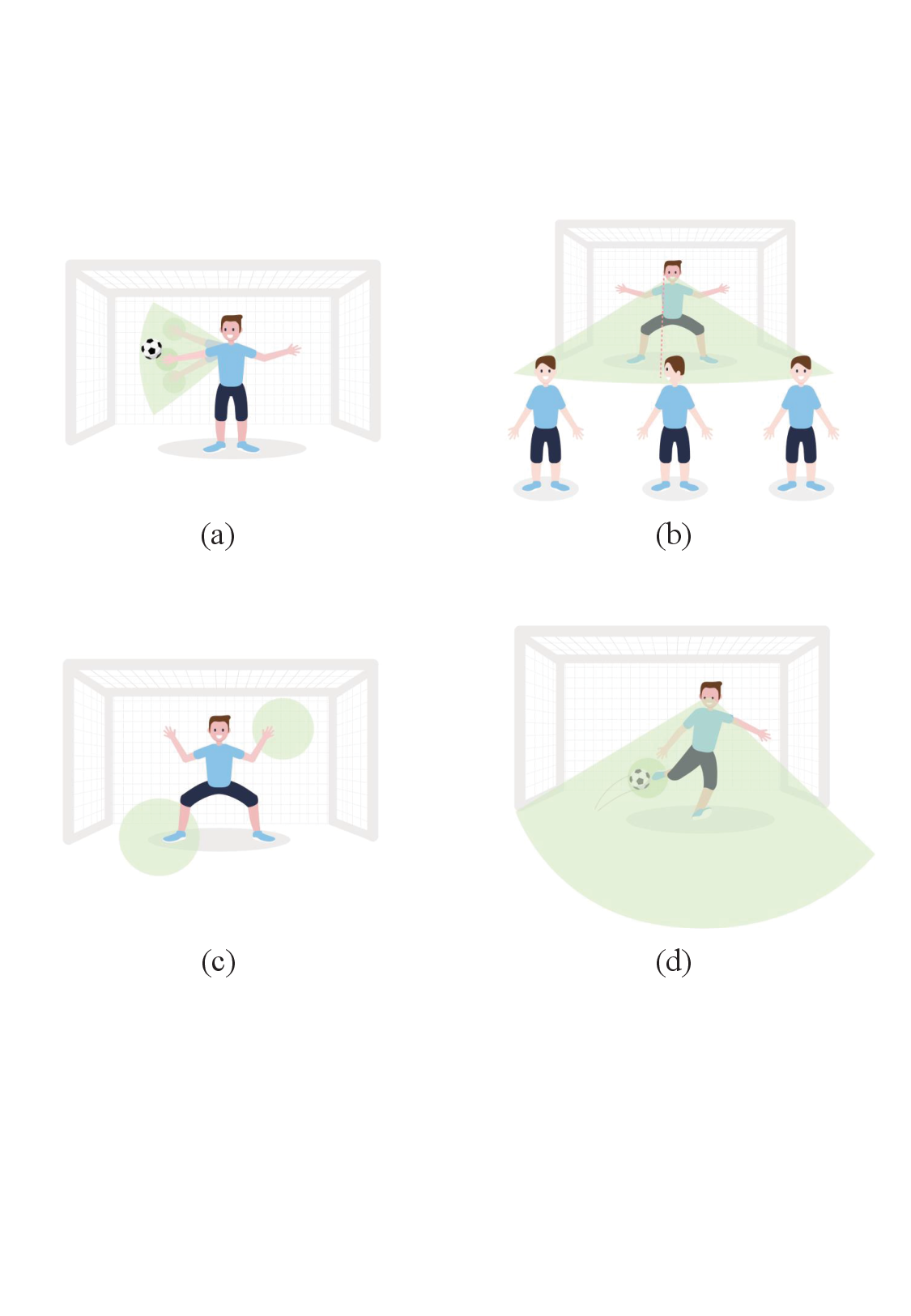}
  \caption{Illustration of hierarchical coordination behaviors of our human body: examples of a goalkeeper in playing football.}\label{fig3}
\end{figure}

An example of playing football is shown in Fig.~\ref{fig3}. The goalkeeper coordinates his eyes, mouth, arms, hands, and feet in different ways according to the varying requirements. Particularly, we can observe the hierarchical coordination between arm and hand in Fig.~\ref{fig3}-(a), where the coverage range of arm is large and that of hand is relatively small. To stop a coming ball, the goalkeeper first moves his arms to a proper position, and then uses his hands to accurately grab the ball. We regard this behavior as arm-hand coordination.
We see the hierarchical coordination between shout (one-to-many broadcast) and whisper (one-to-one communication) in Fig.~\ref{fig3}-(b). In a placement shot defense case, the goalkeeper
first shouts to all his teammates for a general defensive strategy, and then whispers to some teammates, e.g., the central defender in Fig.~\ref{fig3}-(b), for fine adjustment. We regard this behavior as broadcast-communication coordination. The hierarchical coordination between hands and feet is illustrated in Fig.~\ref{fig3}-(c). The goalkeeper may use both his hands and feet to stop a coming ball, depending on the position of the ball. We call it hand-foot coordination. In Fig.~\ref{fig3}-(d), we show the hierarchical coordination between eyes and foot. In a defensive round, after having successfully caught the ball, the goalkeeper first watches the position of his teammates and then kicks the ball towards the desired direction to start a new strike. We regard this behavior as eye-foot coordination.

\begin{figure}[!t]
  \centering
  \includegraphics[width=3.4 in]{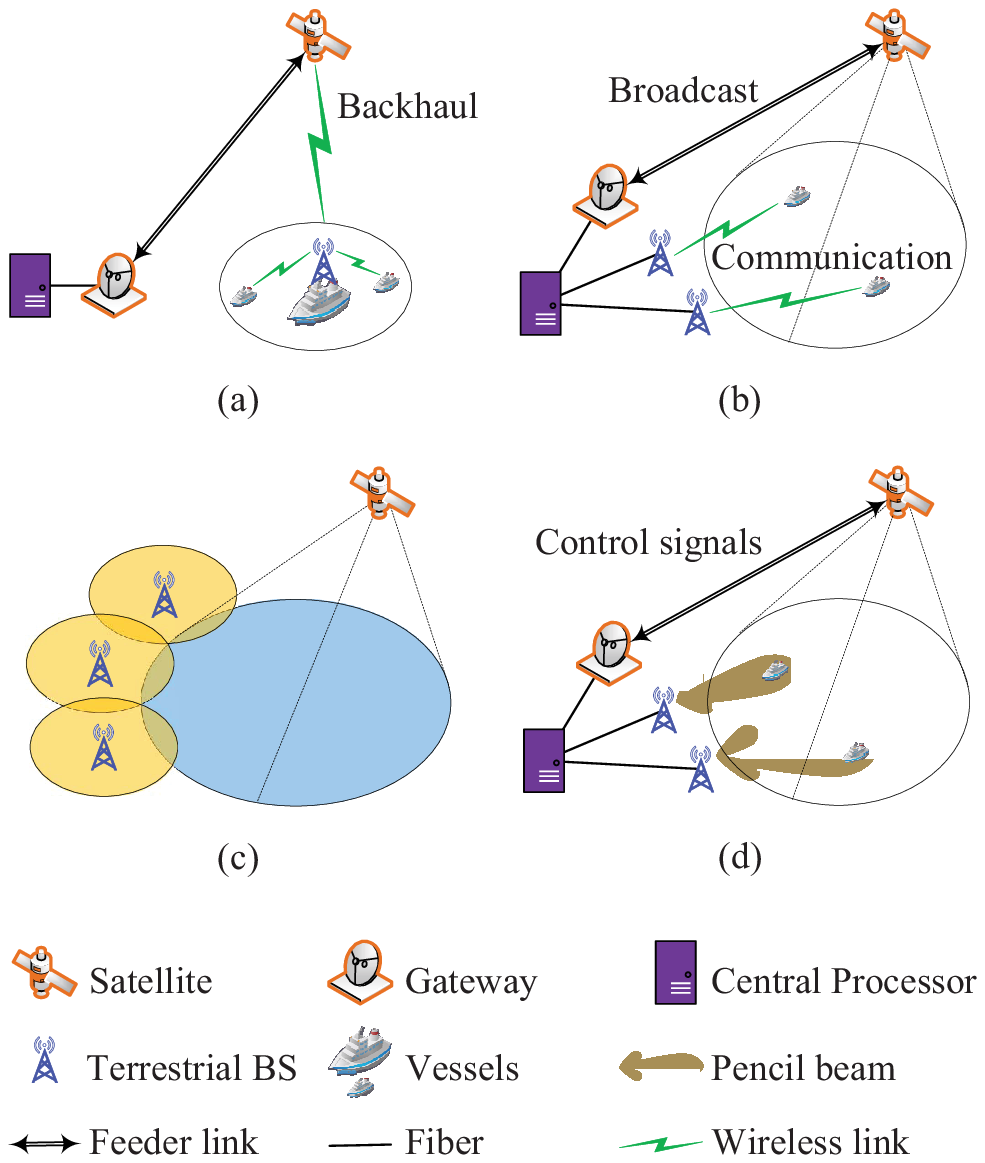}
  \caption{Illustration of four basic structures from a satellite-terrestrial integration perspective.}\label{fig4}
\end{figure}

\begin{figure*}[!t]
  \centering
  \includegraphics[width=6.8 in]{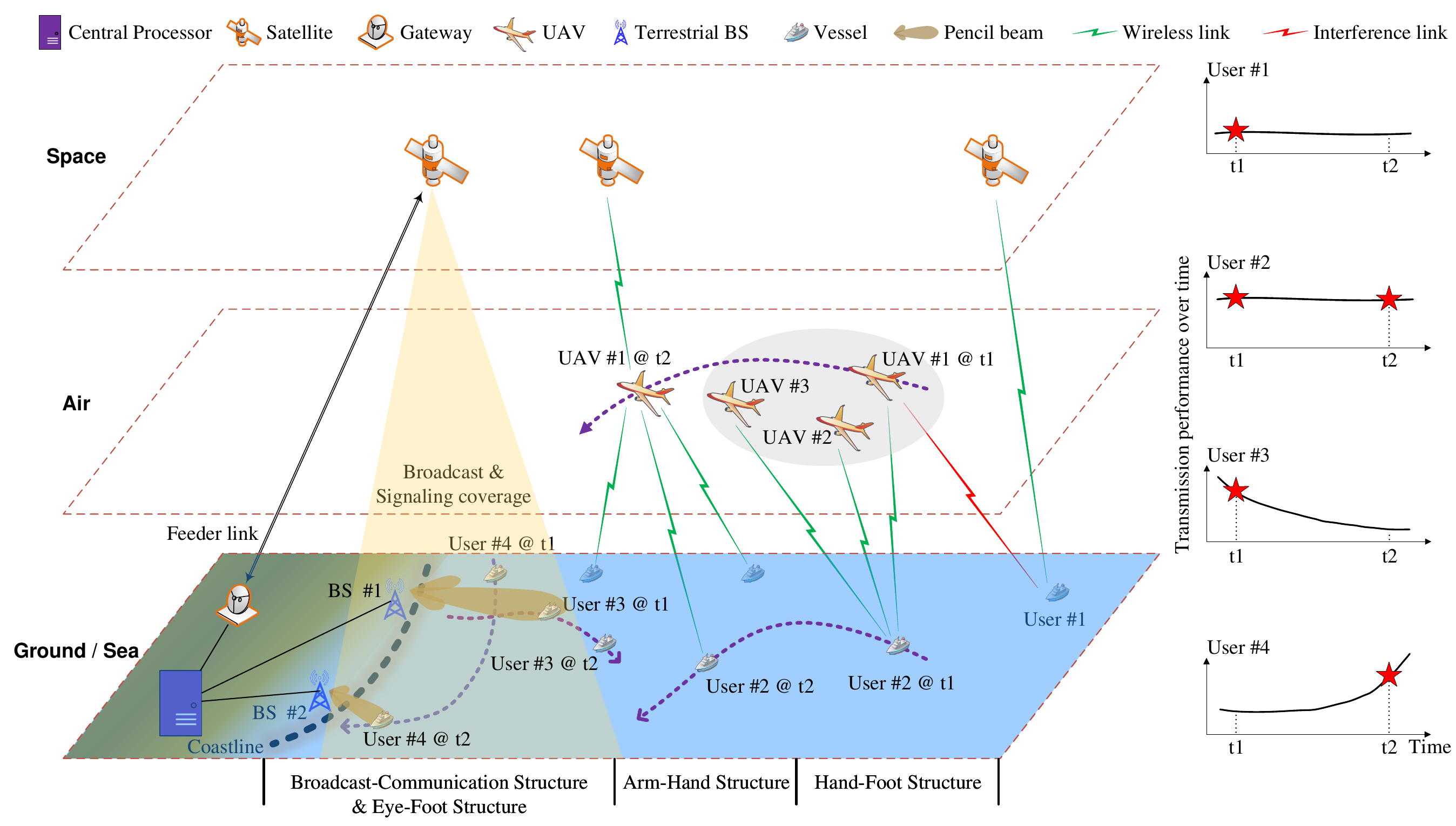}
  \caption{Illustration of the process-oriented on-demand coverage, where the purple dashed lines with arrow denote either the shipping lane of vessels or the trajectory of UAVs.}\label{fig5}
\end{figure*}

We mimic the coordination behaviors of a goalkeeper and accordingly propose four basic structures as shown in Fig.~\ref{fig4}, where the maritime scenario is depicted as an example. We call them arm-hand structure, broadcast-communication structure, hand-foot structure, and eye-foot structure. In the following, we describe them in detail.

\textbf{Arm-Hand Structure.} As shown in Fig.~\ref{fig4}-(a), the large vessel equipped with a high-gain antenna element may receive broadband services from the satellite, whereas the small vessel cannot due to low-gain antennas. In this case, a terrestrial-like mobile BS can be deployed at the large vessel, and serves the surrounding small vessels, using the satellite link as backhaul. This mimics the arm-hand coordination as shown in Fig.~\ref{fig3}-(a).

\textbf{Broadcast-Communication Structure.} As shown in Fig.~\ref{fig4}-(b), the weather report or sea condition information can be broadcasted using the satellite, leveraging its wide-area coverage advantage. Whereas, private data can be transmitted using the terrestrial BS. This mimics the broadcast-communication coordination as shown in Fig.~\ref{fig3}-(b). Its goal is to save communication resources by assigning broadcast services to the satellite system, avoiding repeated transmissions of the same information.

\textbf{Hand-Foot Structure.} As shown in Fig.~\ref{fig4}-(c), all engaged parts covering different areas are used to directly communicate with mobile users, for an extended coverage area. This mimics the hand-foot coordination as shown in Fig.~\ref{fig3}-(c), which is the most straightforward way of integration. However, the edge performance is a crucial issue to be handled.

\textbf{Eye-Foot Structure.} As shown in Fig.~\ref{fig4}-(d), the satellite collects
information of sparsely distributed vessels, e.g., their positions and quality of service requirements, based on which terrestrial BSs adopt accurate pencil beams to serve these users, leaving control and signaling to the satellite. This mimics the eye-foot coordination as shown in Fig.~\ref{fig3}-(d). It is the key methodology to reform the cellular architecture, achieving greatly improved broadband coverage efficiency.

Note that similar integration methods may also apply to the case between satellite and UAV, or between UAV and terrestrial BS. It depends on the specific application settings. Moreover, it is beneficial to treat these basic structures, instead of traditional heterogeneous space/air/ground links, as basic elements in the optimization of an integrated network, so as to decompose the complexity to some extent.

\section{Process-Oriented On-Demand Coverage}
Different from the traditional cellular-based coverage method, on-demand coverage allows for blind zones if there is no user at that time. The system resource is concentrated for online users, thus achieving a higher network efficiency. Under this new framework, the aforementioned basic structures should be elaborately orchestrated according to the user demand. An example can be found in Fig.~\ref{fig5}, where
the roles of UAV \#1 and the satellite are changed from t1 to t2. At t1, a hand-foot structure is formed by UAV \#1 and the satellite, to serve user \#1 and user \#2, where UAV \#2 and UAV \#3 also join to form a UAV swarm~\cite{r8, r12} for carefully mitigating the leakage interference. At t2, an arm-hand structure is formed, where the satellite provides backhaul for the UAV \#1-mounted BS, to serve user \#2. 

As shown in Fig.~\ref{fig5}, if we elaborately exploit the moving process information of both users and UAVs, a larger coverage gain can be achieved. In the figure, both user \#3 and user \#4 follow fixed shipping lanes. Thus, we may account for this process information in coverage optimization, given the delay tolerance of services. The result is quite different from the traditional cellular architecture~\cite{r13}. At t1, although user \#4 is geographically closer to terrestrial BS \#1 than user \#3, terrestrial BS \#1 chooses to allocate the pencil beam to user \#3 under the eye-foot structure, where the satellite provides signaling coverage to guide the pencil beam. This is optimal from the whole-network optimization perspective, as it can be seen in Fig.~\ref{fig5} that user \#4 may enjoy high-performance communications at t2 by terrestrial BS \#2. However, user \#3 goes more and more far away from the terrestrial infrastructures, and t1 is the best moment for it to enjoy broadband serve. In addition to providing signaling coverage, the satellite could also broadcast some public information, forming the broadcast-communication structure, as shown in the figure. 
Thanks to such 
process-oriented on-demand coverage optimization, the achieved transmission performance of user \#2, user \#3, and user \#4 are all satisfactory, as shown in the figure. Being served by the satellite only, thus without the gain of space-air-ground network integration, the performance of user \#1 is the worst. If we use the process information of both UAVs and users, an accompanying coverage can also be achieved as shown in the figure, where UAV \#1 moves accompanying with user \#2, so as to offer a high-performance on-demand service~\cite{r14}.

Due to space limitation, we do not give all details about these examples. One may find both theoretical analysis and simulation results in~\cite{r8,r12,r13,r14}. One thing worth noting is the type of CSI used for process-oriented optimization. Traditionally, the instantaneous CSI is adopted for state-oriented optimization, leading to a variety of optimization methods and tools. For process-oriented optimization, it is impossible to acquire instantaneous CSI. For example, it is impractical to get the CSI between UAV \#1 and users of t2 at t1, because the CSI between t1 and t2 may be uncorrelated, i.e., t2-t1 is much larger than the coherent time for traditional CSI prediction methods. We thus suggest to use the location-dependent large-scale CSI~\cite{r15}, which can be obtained by learning from historical data together with electromagnetic calculations~\cite{r8,r12,r13,r14}.

\begin{figure}[!t]
  \centering
  \includegraphics[width=3.2 in]{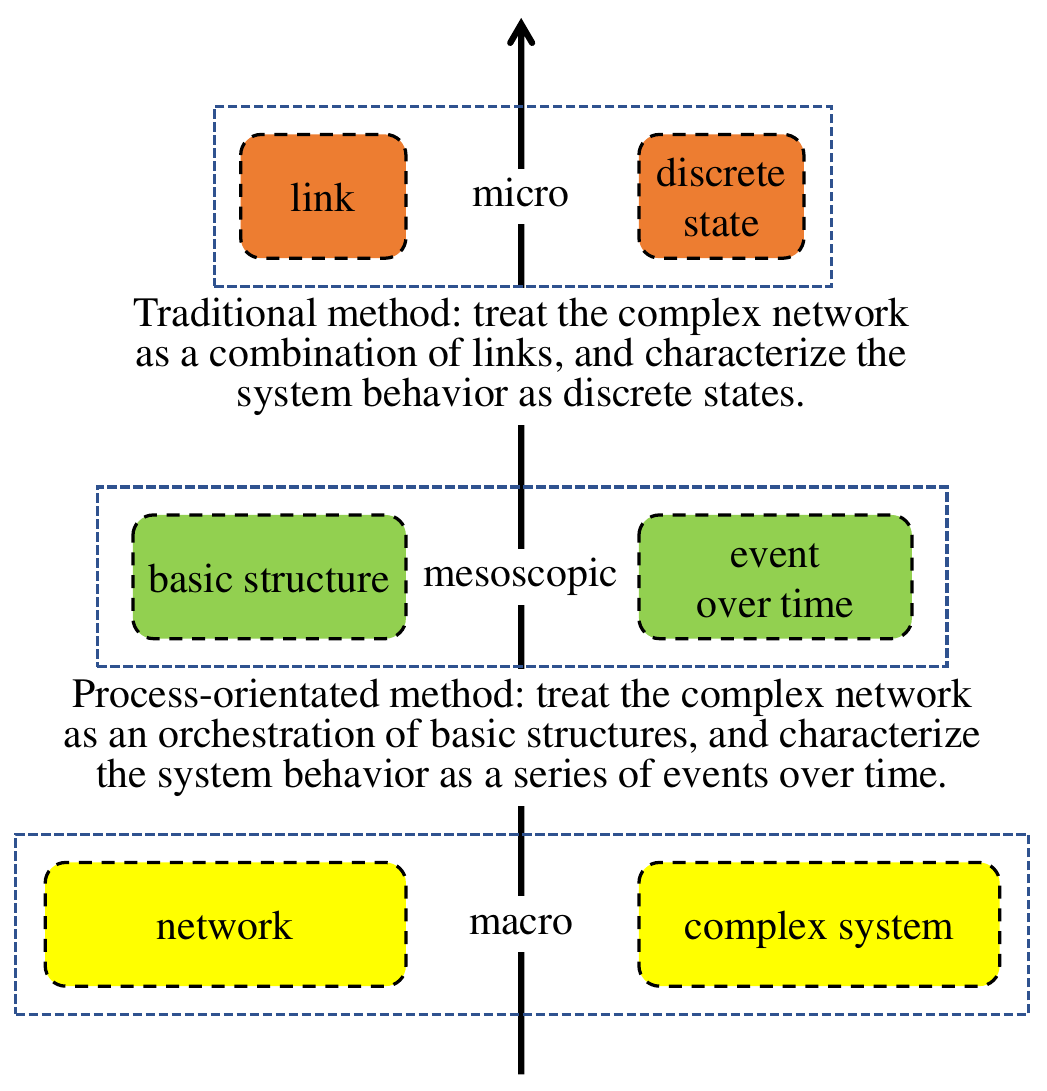}
  \caption{Comparison between the proposed method and traditional one.}\label{fig6}
\end{figure}

We summarize the proposed method in Fig.~\ref{fig6}. Traditionally, the network is regarded as a combination of links, and the behavior of system is usually characterized by discrete states. This leads to a variety of  
state-oriented and link-based optimization approaches,
which are efficient for the cellular architecture due to its decomposability and stationary characteristics. For the integrated satellite-UAV-terrestrial network, both the heterogeneity of space/air/ground links and the dynamics of system behavior will challenge traditional methods.
We propose to treat the complex network as an orchestration of basic structures and characterize the system behavior as a series of events over time. Consequently, we present the process-oriented on-demand coverage optimization method, which actually solves the problem at a mesoscopic scale, the middle one between conventional micro and macro scales.

\section{Open Issues}
In the section, we briefly discuss some open issues arising from the proposed method to stimulate more researches.

Firstly, new performance metrics for the proposed basic structures and their orchestrations should be established. This might be quite different from the traditional Shannon capacity, because both the user distribution and service type should be taken into account. Nevertheless, the new metric
should be as general as possible for ease of implementations, rather than being quite case-sensitive.

Secondly, a new network framework should be presented to adapt to the process-oriented optimization approach. If there is a feedback in the process, the optimization can be accumulated, which may even produce a knowledge base after learning. This knowledge base would be quite useful, as it may empower an incremental optimization, leading to a much more stable and smarter system. Therefore, the new network framework should be knowledge-driven and even human-like. To be much more expected, it breaks the 10-year one-cycle upgrading of wireless communication networks, providing
a new pathway for continuable progressive evolutions. Instead of thinking only from the communication perspective, one may have to rethink 6G network framework from a whole-chain information engineering perspective that takes into account all segments of information processing, including sensing, transmission, storage, computing and control. In this case, this framework should be designed with sufficient intelligence so as to match both the upstream and downstream segments of information engineering.

Thirdly, the frequent shift between different basic structures will inevitably lead to the data exchange issue. How to guarantee the accuracy and integrity of the data is an open issue. The emerging blockchain technology can be introduced, which may also help to establish an integrated space-air-ground spectrum sharing regime for alleviating the spectrum scarity problem. For users, both privacy and ownership of data should be protected in the 6G era. For remote manipulation or intelligent robots, the accuracy and integrity of controlling signals should be guaranteed. In a nutshell, 6G ought to be more secure in a systematic manner.

\section{Conclusions}
In this article, we have investigated the shortcomings of current 5G networks and envisioned new use cases for 6G IoT. To tackle the complexity in integrated satellite-UAV-terrestrial networks, we have proposed to create and exploit network hierarchies from a systematic thinking. Four basic structures have been established by learning from synergies in our human body. Consequently, by regarding the complex network as an orchestration of basic structures and characterizing the system behavior as a series of events over time, we have proposed the process-oriented on-demand coverage optimization method. Our method can solve the problem at a mesoscopic scale and has shed lights on the systematic thinking-empowered network optimization in the 6G era.

\end{document}